Dollarized Economies in Latin America. An Inflationary Analysis of Pre, During and

Post Pandemic.

Authors:

Joseph Ariel Tello Carvache

Jorge Alejandro Moncayo Correa

MBA. Carlos sempertegui

December 2023

Abstract

Given the hyperinflation that most of the Latin American countries suffered in the 90's and their decision towards adopting dollarization and in most cases keeping their own currency, this paper analyzes the effectiveness of dollarization as a protective mechanism against economic disruptions in Latin American countries. It assesses the context that led Latin American dollarized countries to dollarize and analyzes CPI, GDP, and the poverty rates pre, during, and post-pandemic in Latin American countries, considering those that are dollarized and those that are not, and evaluating its relation to the US. Interviews were carried out with experts in the field. It assesses the advantages and disadvantages of dollarization regarding global crises. The data was compared and analyzed to check if there were patterns that support the paper's objective which is that dollarization might serve as a protective mechanism against economic disruption. It was found that dollarization protects the economy against inflation, however, it does not fully protect the economy when considering economic performance and poverty. In conclusion, this research concludes that dollarization does not completely serve as a protective mechanism against economic disruptions; nonetheless, it found that a bigger role is played by domestic policies and government action.

Introduction

"Those who cannot remember the past are condemned to repeat it" (Santayana, 1905). On March 8, 1999, Ecuadorians awoke to discovered that their bank deposits were freeze in a period called bank holiday. In fact, in the late 1990's Ecuador was going through a strong inflationary phase due to several factors such as El Niño phenomenon, foreign debt, political issues, and so on. Ecuador reached an inflation of more than 100%, leading the government to take drastic measures for ensuring economic stability in the long run, considering the banking failures that were happening (Ceja, 2016; Reed, 2020). Another example is El Salvador, after going through a crude civil war, the inflation reached almost 18%, despite inflation stabilizing in the late 1990's, interest rates were the highest in the region, making the average Salvadorean almost impossible to afford a loan (Kopar, 2018). Additionally, In the early 2001, Argentina faced a dramatic economic crisis product of the excessive government expenditure, resources mismanagement, and the fall of the peso's convertibility into dollars. Notably, this led to an inflationary phase of almost 40% and making Argentinians to loss a great part of their deposits due to the new floatable exchange rate (Cortés Conde, 2003; Instituto Nacional de Estadística y Censos, 2023). On the other hand, there is a Latin American country that has not suffered hyperinflation or a significant economic turmoil as others Latin American countries. In fact, Panama has the more stable economy in the region; however, it is also susceptible to economic and social problems, especially when you have a booming economy where part of the population is living in poverty (Archibold, 2013; EFE, 2022).

As a matter of fact, inflation plays a significant role in every economy around the world, especially because it directly affects the purchasing power of the individual; hence, affecting its wellbeing in several manners. However, the problem starts when inflation levels are not stable, meaning that they can rise dramatically from one period to the other. For example, high inflationary rates such as what happened in the example that were described above, affected the individual's capacity to acquire goods and maintain proper living conditions such as acquiring a complete basic basket and living a decent live (Floyd, 2023).

Notably, during an economic turmoil central banks use tools such as expansionary monetary policies for stimulating economic growth and recovery (Suarez, 2022). Nonetheless, it is worth noting that by increasing the money supply the inner value of the currency is affected, meaning that inflation occurs. The extent to which inflation occurs is determined by how much money has been injected into the economy. In the case of Colombia, they were aiming to increase interest rates 100 basis point from 4% to 5% to reduce inflation for 2022 and 2023. In fact, their projected inflation was 6.4% and 3.8% respectively. However, their inflation ended up being 13.2% for 2022 regardless of the adopted measures as a consequence of an expansionary measured adopted during the COVID-19 pandemic (DANE, 2023; Focus Economics, 2022).

Another example that is worth considering is Argentina. On 2018 the country was experiencing an average of 33,8% inflation, and reaching a 53,8% inflation for December 2019, having its peak inflation on May of the same year reaching 57,3%. Surprisingly, for 2020 the country experienced a declining rate of around 1 to 2% each month, ending at 35,8% for December 2020, stabilizing again with respect to 2018. Argentina had adopted an exchange rate regime which performed daily micro-devaluations to avoid exchange rate lag, but inflation had accelerated in the last months (Gaspar, 2023). From this point forward, inflation started increasing at an uncontrollable rate, with an average of 48,1% in 2021 and 70,7% in 2022, ending the year reaching a peak of 94,8% inflation, and continues increasing from there with a forecasted increase of over 100% inflation. (INDEC: Instituto Nacional de Estadística y Censos de La República Argentina, 2023)

As a measure of controlling inflation, Latin-American countries have pursued dollarization; the adoption of the U.S. dollar as an addition to, or substitution of the domestic currency of another country, recognizing the U.S. dollar, official United States, and designated world reserves' currency, as a medium of exchange either formally or informally, mainly due to its relatively low inflation rate. While there are both benefits and drawbacks, it has become a considered option in recent decades for Latin American countries. Especially due to hyperinflation or economic instability, or diverse factors which have a strong impact on the domestic currency, becoming unstable and losing its usefulness as a medium of exchange for market transactions. By adopting the U.S. dollar, it is expected an enhanced monetary and economic stability resulting from the adoption of the currency, but the action of turning to the U.S. dollar necessarily involves a loss of economic autonomy over monetary policy.

This paper will analyze the extent to which dollarization serves as a protective mechanism against economic disruptions in Latin American countries. For doing so, it is paramount to consider the following points. First, assessing the context that led Latin American dollarized countries to dollarized. This will provide a background for understanding the effects of dollarization in dollarized Latin American countries. Second, gathering and analyzing inflation rates, gross domestic product "GDP", and the poverty rate for pre, during, and post pandemic in Latin American countries, considering those that are dollarized and those that are not, followed by the evaluation of its relation to the US, providing valuable patterns that the selected countries followed before, during, and after an economic disruption (COVID-19). Third, analyzing advantages and disadvantages of dollarization regarding global crises, focusing on economic stability, economic development, and social implications contributing with valuable information when stating the extent in which dollarization might serve as a protective mechanism against global disruption. In short, the objective of this paper is to present the potential that dollarization has for Latin American developing countries when considering economic stability.

## Methodology

Notably as this paper is aiming to present the extent to which dollarization serves as a protective mechanism for economic disruptions in Latin American countries, qualitative and quantitative analysis will be used for the opportunity for analyzing and assessing economic indicators, performance, history of dollarization in Latin American countries, and compare them before, during, and after the pandemic.

As a matter of fact, qualitative analysis refers to the analysis of information that cannot be counted and that must be analyzed and reflected by the individual (Tech Target, n.d.). In this case the information that is going to be employed is related to the historical context regarding dollarized Latin American economies (Ecuador, El Salvador, and Panama) and what led them to opt for dollarization. Additionally, Interviews will be used to present the context in which nations decided to be dollarized, understand the spectrum in which it was carried out, and why it was implemented. In addition to this, the central topics of the interviews will be benefits and drawbacks of dollarization for Ecuador, what determines the success of dollarization, and their opinions about the economic landscape of Ecuador in the post-pandemic era. The interviews will be carried out to experts such as the Ecuadorian Economist Alberto Dahik Garzozi, former vice president of Ecuador and an expert in economic policies. To distinguished professors such as Oswaldo Patiño Vega and Julio Vasquez Moreira specialized in economics and finance respectively, both that have experienced in dollarization. Interviews will be conducted in a semi-structured format, meaning that there will be room for discussion as well as flexibility on the topics, providing in this manner the opportunity to get their perspectives as clear as possible. Questions can be seen on annex 1. Without any doubt, it will provide valuable insights when considering and analyzing the extent to which dollarization has served as a tool that ensures economic stability when economic turmoil happens.

According to Sebastian Taylor, quantitative analysis is referred as "the process of collecting and evaluating measurable and verifiable data" (Taylor, 2020). Regarding the analysis to be directed, the considered information to be employed for the research are economic indicators such as the Gross Domestic Product (GDP), Consumer Price Index (CPI) as a measurement for inflation, and Poverty rates as a measurement for society to be related to the economic indicators. This research focuses on pre, during, and post pandemic timeline, it is essential to do an analysis from the year 2018 up to 2022. These indicators will provide specific information about the country's situation and allow an efficient comparison. In fact, with these methods it will be possible to analyze the effectiveness of dollarization on the chosen countries (Ecuador, El Salvador, and Panama) with respect to those non-dollarized (Colombia and Argentina) at ensuring economic stability during economic disruptions and considering the US as a reference. Along with the economic factors, it will be compared and relate it to the poverty rate of the same country, looking if there is a trend between these indicators' performances. Quantitative data of the Covid-19 will be used among the already mentioned indicators, as mostly the information needed from this is qualitative. It will be used statistical techniques to describe the tendency, time trend, and variation between 2018 to 2022 using Microsoft excel and IBM SPSS as analysis tools.

Body

As a matter of fact, an economic crisis refers to the contraction, depression, or recession of the economy. Most of the time, crisis happens in the blink of an eye due to resources mismanagement, conflicts, inappropriate usage of monetary policies, and so on (Chen, 2021). On the other hand, when talking about economic crisis, is essential to consider inflation, especially when talking about the Latin American context. In fact, inflation refers to the reduction of purchasing power due to the rise of average prices in a nation, affecting in this manner their wellbeing and lifestyle of the individual (Fernando, 2023). When considering the Latin American context, it is paramount to assess the history of dollarized countries and why they decided to dollarize.

Ecuador in the 1990's went through several events such as the El Niño phenomenon, the drop of oil prices, and political instability. Factors that lead the government to implement dollarization in the nation. First, in the case of a natural disaster such as "El Niño" phenomenon of 1997 is essential to consider that negative economic effects are originated from the loss of capital and infrastructure assets. In fact, according to the United Nation, the speed of recovery after a natural disaster is one of the most important things when considering economic recovery and social well-being (United Nations, 2016). For that reason, governments recur to debt with the purpose of providing the necessary resources for an adequate economic and social recovery. On the other hand, when talking about the private sector, loans cannot be set aside, especially when borrowers cannot pay back their loans. This means that given the mentioned scenario, the banking sector was severely affected and Ecuador faced a loss of $2,882 million dollars (Basantes, 2020; El Universo, 2023; García, 2013). Second, in 1998 the oil price dropped from $45 to almost $7 per barrel. Considering that Ecuador is heavily dependent on oil, the drop of oil prices severely affected the economy specially in the fiscal aspect and budget. In other words, according to the central bank of Ecuador, the budget for 1998 was made using a forecasted average price of $16; however, the real price ended up being an average of $9.19 per barrel which significantly affected the economy (Banco Central del Ecuador, 2000). Third, political instability played an important role in debilitating the Ecuadorian economy, in fact, from 1996 to 1999 Ecuador had 4 presidents due to destitutions and social unsatisfaction. This clearly impacted investor's perception and Ecuadorians, especially when bank runs, and the

uncertainty of the financial system were suggesting a potential collapse of the system. Consequently, Ecuadorians started to withdraw money from their savings accounts and investors were taking out liquidity from many industries. Hence, debilitating the financial system up to a point where the lack of liquidity and instability of the financial sector drove the government to freeze savings accounts and to opt for dollarization as a way for improving economic trust and financial stability. (García, 2013)

El Salvador, like Ecuador, has faced a series of economic challenges during the 1990's which ultimately led to the adoption of the U.S. Dollar is 2001. Relevant events that affected and contributed to the alternative of dollarizing the country include the El Niño Phenomenon during 1998. This event caused heavy damage to the country's infrastructure and had a great affection on the agricultural sector, which led to a significant economic loss during this period. (El Universo, 2023) The government's quick response to these events was to make heavy borrowings to finance reconstruction efforts, the same which later would be reflected critically on the country's finances. Alongside, the country's main export crop, coffee, experienced a sharp decline in global coffee prices in the late 1990's. (Brown, 2014) In the same period, El Salvador went through a significant experience of political upheaval, times in which the government suffered many changes in short periods of time, and social unrest. (FMLN, 2006; González, 1999) The political instability had a strong impact on the economy of the country and on the investors' confidence, making it hard for the country to be attractive for foreign investment, hence impeding economic growth. The Salvadoran government, in response to the country's economic and political situation decided to implement dollarization. This method was considered and implemented expecting it to be a stabilizing mechanism to the economy in firsthand, and consequently be able to recover the confidence of foreign investors for them to keep investing in the country, as there would be less threats economically speaking for them.

As a quick response from this transition event, El Salvador was able to effectively eliminate exchange rate risk and reduce inflation.

Panama was a country that lacked a stable national currency and relied on a bundle of Colombian pesos, other foreign coins and even private local businesses private banknotes. This system led to several challenges like inflation, currency fluctuations and a lack of trust in the financial system. The fact that the country did not have an official currency was limiting the international trade participation and made it difficult for the Panamanian economy to develop. The country was exposed to external shocks, from the use of foreign currencies rather than a national currency, and potential fraud from the proliferation of private banknotes.

The need for monetary reform was becoming increasingly apparent by the early 1900's, yet the Colombian government remained hesitant to allow Panama the establishment of its own currency. Three years later, after a realization on the country's lack of progress, Panama separated from Colombia and were left with no currency, and under a higly risky inflation on the economy. By the following year, in 1904, the Panamanian government decided to adopt the U.S. dollar as their legal tender, and constitute the balboa as its official currency, yet this last one exists only as coins and circulate along with U.S. dollar bills, creating a dual currency system.

Notably, monetary policy is one of the most important tools that a central bank has for controlling the money supply as well as promoting economic growth. In other words, as it controls the money supply, it has the capacity to manage inflation, and depending on the circumstances, it can promote economic growth and development. Nonetheless, there are advantages and disadvantages of having such a tool. Monetary policy is divided into contractionary or expansionary and its use depends on the result that the central bank is aiming to achieve. In the case of inflation, central banks employ contractionary policies to reduce the money supply; hence, reducing inflation to stable figures. Additionally, if a country desires to be more competitive for exportations or needs to boost its economy, the central bank will employ an expansionary policy, meaning that it will decrease the value of its currency when compared against others, and it will decrease interest rates in order to boost the economy (Brock, 2023). On the other hand, according to the Economist Oswaldo Patiño, monetary policy is important; however, the problem starts when governments do not have monetary discipline, meaning that they employ such tactics without considering the potential effects that they might have on the economy. For example, in the case of Ecuador, they indiscriminately increased its money supply due to excessive fiscal expenditures and mismanagement of external events such as "El Niño" Phenomenon, the drop of oil prices, and political instability. Events that encouraged the central bank to indiscriminately employ expansionary policies without proper planning. Hence, causing inflation to go from 25% to almost 90% in a matter of a year and provoking the financial crisis of 1999-2000 (García, 2013; INEC, 2023b; Patiño, 2023). Another interesting example of the abuse of monetary policy is Argentina. From 1991 to 2001 the public debt increased 129%, affecting in this manner the investors and public trust on the financial system. Due to the lack of trust in the system, international entities restricted credits for Argentina, forcing them to recur to expansionary monetary policies. It is worth noting that the fiscal expenditure was excessive considering the given economic conditions such as the flight of capital, investment reduction, distrust on the financial system (Galiani et al., 2003). With excessive fiscal deficit, Argentina decided to stop its peg to the dollar causing the peso to freely float; hence, provoking a sudden inflation of 40% and damaging investor's confidence on the financial system. In short, the given examples present how the lack of monetary discipline and control can lead a country to experience unexpected inflation levels and therefore affecting the citizens due to resources mismanagement (Economia y Desarrollo, 2020).

Without any doubt, dollarized economies do not have the ability to influence or manage its own monetary policy given that they only one who control's it is the Federal Reserve of the United States of America. In other terms, by not having such a tool, they are unable to manage inflation, boost their economy, or manipulate its currency value to be more competitive on the international market. Nonetheless, there are several benefits such as price stability, non exchange rate risk, investors' confidence, low interest rates, and economic policy discipline. It is important to consider that dollarization should be seen as a tool that enhances stability but is up to the country to take advantage of it. In other words, to take advantage of dollarization, countries should focus on developing their industries in the long term instead of trying to compete in the commodities market where they are at a disadvantage when considering exchange rates (Patiño, 2023; Vasquez, 2023).

As a matter of fact, when analyzing dollarization is essential to consider the results of implementing such an economic tool. As it was previously mentioned, Ecuador in 1999 was going through an economic turmoil due to several factors such as political instability, fiscal deficit, resources mismanagement, and so on. Factors that lead inflation to reach almost 100 percent, reducing in this manner Ecuadorian's purchasing power; hence, affecting their well being. On January 9$^{th}$, 2000, former president Jamil Mahuad opted to dollarize the country as a measure for counteract the imminent hyperinflation that was around the corner (CNN, 2023; Meléndez, 2022). Without any doubt, dollarization has served its purpose in Ecuador, in fact according to the world bank; in fact, due to dollarization, the country was able to reduce its inflation from almost 100 percent to 2.7 percent in a matter of 4 years (World Bank, n.d.-b). In addition to this, in the same period, the gpd was improved from almost -4 percent in 1999 to 8.2 percent in 2004 (World Bank, n.d.-a). In other words, considering the impossibility of employing monetary policies, dollarization forced Ecuador to be fiscally disciplined. And, at the same time it has provided several advantages such as a better investment environment considering the inexistence of exchange rate risks, price stability, and the reduction of poverty from 44.8% in 1998 to 27.7% in 2021 (Ortiz, 2023).

Undoubtedly, dollarization has played a key role in stabilizing the Ecuadorian economy; however, it essential to analyze the drawbacks that the Ecuadorian dollarization has had. Notably, dollarization has a dark past in the country when considering the society's suffering due to the exchange rate volatility; in fact, in 1999, 5,000 sucres were the equivalent of 1 dollar; while when dollarization was implemented, it raised to 25,000 sucres per 1 dollar. This means that Ecuadorians savings were practically evaporated overnight affecting in this manner the middle class and retirees (Roura, 2020). Besides this, dollarization significantly affected the Ecuadorian competitivity on international markets due to the impossibility to devaluate its currency against other Latin American competitor's currencies. However, it is worth noting that this impact is not only rooted in dollarization, instead it comes from not having improved the country's productivity since the 2000s. In other terms, before and after dollarization, Ecuador has been solely focused on selling commodities such as oil and commercializing raw materials such as bananas, shrimp, flowers, and so on. Consequently, making Ecuador unfit for competing against countries that have the capacity to devaluate their currency. Apart from this, the lack of monetary policy acts as a brake when considering the country's development, especially when the nation hasn't developed its industry and has mismanaged its resources. Also, unemployment in Ecuador is 3.8 percent, while poverty is 27 percent in the first quarter of 2023. However, it is worth noting that without the existence of a strong industrial sector, it is hard to offer better salaries considering labor force qualifications; hence, making organized crime easier to spread in the country (Álvarez, 2023; Escobar, 2023; INEC, 2023d).

The Salvadorian Colon was replaced by the U.S. dollar by the beginning of 2001. The immediate response to the country was the stabilization of the economy and the high inflation it was experiencing, reducing it by around 2% (Swiston, 2011). El Salvador shares common results at dollarization, as the country also had an increase in investment, due to its increase in attractiveness to foreign investors at using the U.S. dollar; Not only the use of the currency but also the fact that investments were no longer running under the same inflation risk and devaluation of the currency the country was experiencing before (Colantuoni, 2023). Along with this, El Salvador also benefited of lower interest rates, as the banks were no longer exposed to the currency fluctuations risk and were now under the U.S. monetary policy, making it cheaper for businesses to borrow money and stimulating the economy by doing so (Swiston, 2012, p. 7). El Salvador is under the U.S. monetary policy, meaning that the country lost their monetary sovereignty and its monetary policy. The country became unable to set its own interest rates and/or print its money, relying entirely on the U.S. economic situation, which makes it difficult for the country to respond to economic shocks by their own. In addition, the country's exports could become less competitive as the country no longer possesses a currency to devaluate to make exports cheaper and more attractive. This represents a problem in terms of the economy's heavy dependence on exports.

In social aspects, dollarization is accused of increasing inequality of El Salvador (Booth et al., 2018). This is followed up by the difficulty the use of the U.S. dollar brings to the country for the poor people at saving money. When the colón was in circulation, poor people could afford to save money by buying colones, which would then appreciate as inflation fell; However, with the use of U.S. dollars, people were only able to save money by this same, more expensive, currency.

Notably, the first appearance of Covid-19 was on early November 2019; however, the World Health Organization declared it as a pandemic on March 11, 2020 (Rath, 2022). Without any doubt, the pandemic came with significant social and economic challenges such as lockdowns, travel restrictions, poverty, bankruptcies, supply chain disruptions, and others. Factors that lead to the decrease in productivity and the appearance of uncertainty. The crisis had a severe effect on worldwide poverty and inequality. For the first time in a generation, global poverty rose, and disproportionate income losses among disadvantaged people resulted in a substantial increase in inequality between nations. According to the World Bank, in 2020, temporary unemployment was greater in 70% of all nations for workers with just an elementary education (World Bank, 2022). Additionally, it is worth noting that inflation plays an important role during any crisis. In fact, worldwide inflation was in average 2.09 percent before the start of the pandemic; however, in 2021 raised to 3.48 percent and in 2022 it raised sharply to 8.27 percent (Macrotrends, 2023d). Clearly, it is possible to notice that overall inflation did not happen immediately, this happens because despite rescue plans, people were not able to work due to the inexistence of a vaccine. However, once the people started to get vaccinated and therefore going to work, it put a pressure on wages and prices (National Bureau of Economic Research, 2023).

Without any doubt, during crisis governments must intervene and during Covid-19 It was not the exception. In fact, to manage the economic and social situation in United States, the government decided to offer a stimulus package of $5 trillion. It is worth noting that during the 2008 economic crisis they issue a stimulus package of $787 billion. In other words, Covid encouraged the government to implement extreme measures due to the situation's severity. Apart from that, during Covid-19 lockdown, there was a rise of savings because the society was not expending as much due to uncertainty. Notably, in winter 2020 bank savings went from $13.5 trillion to $15.6; nonetheless, by the end of 2021 they reach $18 trillion dollars. The rise of savings, the aggressive stimulus package, supply chain disruption, the elimination of reserve requirements for US banks, and the rise of commodity prices, are factors that

contribute to inflation in the 2022 once the society was released from confinement, henge putting pressure on wages and prices (McKitrick, 2022; Pettinger, 2022).

Importantly, dollarized economies as its name implies, are tightened to the US economy, meaning that its action over monetary policy affects dollarized economies because it affects the overall supply of dollars. For example, during the pandemic, the US government decided to put interest rates at 0 percent at the beginning of the pandemic (2020); hence, making it cheaper to get a loan. In addition to this, Ecuador interest rates drop from 9 percent in 2021 to almost 6.6 percent in 2022. In the case of Panama, they dropped from 1.5% at the beginning of the pandemic (2020) to almost 0 percent in 2021 and early 2022. And, in the case of El Salvador, interest rates drop from 4.25 percent at the beginning of the pandemic to 3.75 percent in 2021. In this case is possible to notice that dollarized economies apparently followed the US behavior towards interest rates. Nonetheless, it is worth considering that it was not immediately, and its difference is considerable when compared among them. In other words, despite the US setting interest rates at 0 percent, only Panama was able to follow almost the same behavior. This happens due to the country's risk measures. In other words, as a country possess a higher country risk, lenders demand a premium for taking greater risks, as it is lending in Ecuador or El Salvador (Banco Central del Ecuador, 2023; Central Reserve Bank of El Salvador, 2023; FED, 2023b; IMF, 2023). In other words, by making loans more accessible, inflation tends to increase; additionally, during covid the US used rescue packages that lead to the increase of commodity prices (World Economic Forum, 2022). In short, dollarized economies tend to reflect a similar behavior of the US respectively. Nonetheless, factors such as country risk that considers political instability, economic policies, the legal system, and creditworthiness, prevent these economies from mirroring the US.

Figure 1

Correlation analysis of Inflation by M1 in Latin American Countries and

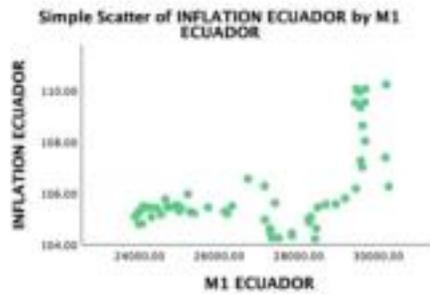
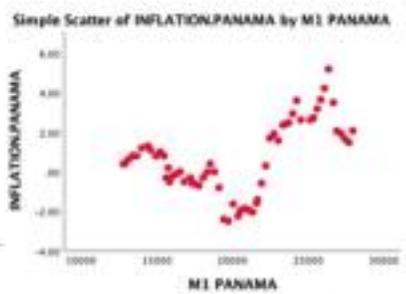

USA

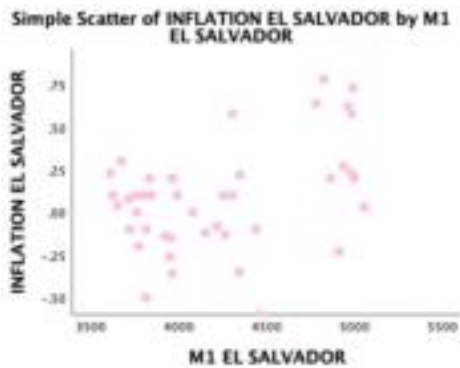
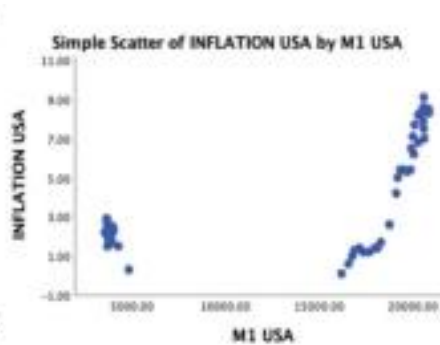

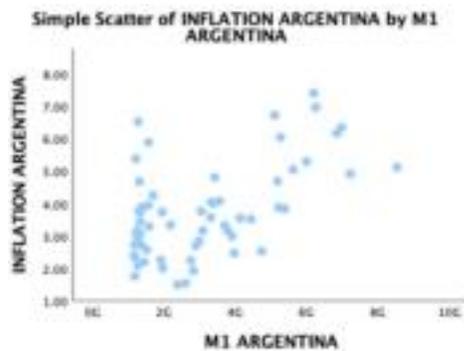
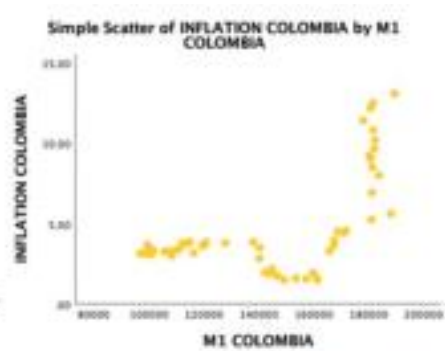

Note. Correlation and relation between M1 and Inflation in dollarized Latin American Countries, non-dollarized Latin American Countries, and the USA. Data

retrieved from (Banco Central de la República de Argentina, 2023; Banco Central de Reserva de El Salvador, n.d.; Banco de la República, 2023; Central Bank of Ecuador, 2023; FED, 2023a; INEC, 2023a; World Bank, 2023b).

Notably, monetary aggregates consider the overall supply of money in an economy. In the case of M1, a component of the monetary aggregates refers to the most liquid form of money including cash, demand deposits, and other assets that can be converted into cash. In other words, M1 presents the overall supply of money in a country (Kelly & Costagliola, 2021).

When analyzing figure 1 it is possible to notice that when M1 increases inflation also tends to increase. In fact, Panama has a moderate correlation of 0.495 and P=0.000 which means that the higher M1, the higher the inflation. In the case of El Salvador, it has a moderate correlation of 0.412 and P= 0.005 which says that the greater M1 the greater the inflation. In the case of Ecuador, it has a considerable correlation of 0.572 and P=0.000 which means that the greater the M1 the greater the inflation. In the case of USA, they have a considerable correlation of 0.656 which is the greatest in the sample and has P=0.000; this means that the higher M1 the higher the inflation. In the case of Argentina, it has a considerable correlation of 0.552 and P=0.000 this suggest that the greater M1, the greater the inflation. In the case of Colombia, they have a considerable correlation of 0.605 and P=0.000 this means that the greater the increase of M1, the greater the rise of inflation.

Figure 2

Inflation Rates in Latin American Dollarized countries and the USA.

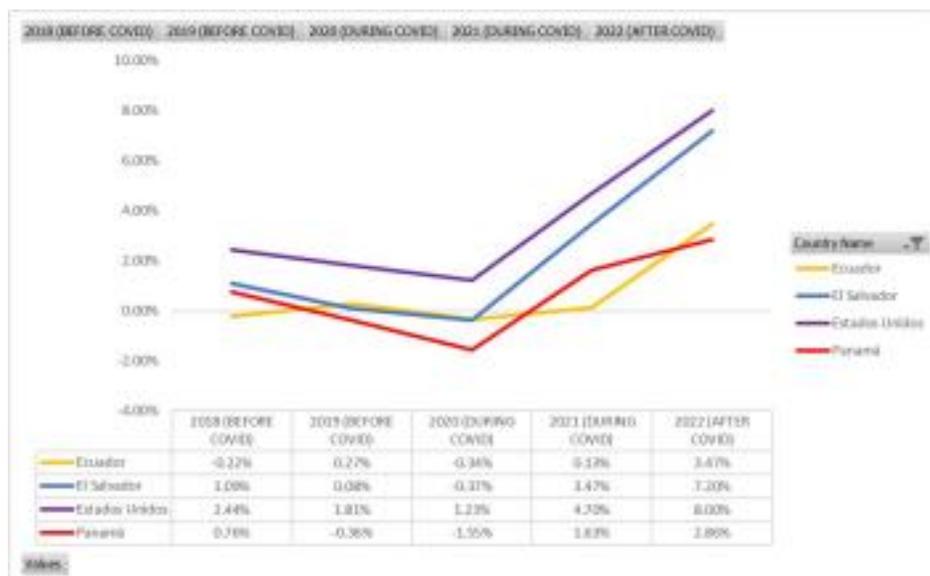

Note. Inflation rates of dollarized Latin American nations and the USA. Data retrieved form (World Bank, 2023b)

Figure 2 shows the inflation evolution in Ecuador, El Salvador, United States and Panama for the period of 2018 to 2022. It is evident that all four countries experimented different inflation trajectories during the COVID-19 pandemic. For Ecuador, there was a negative inflation, or deflation, between years 2018 and 2020, but recovering by 2022 reaching up to 3,47% thanks to the post-pandemic economic recovery. For El Salvador, there was a moderate inflation in 2018, but contracted in 2020, experiencing a deflation under the COVID

19 pandemic circumstances. By 2022, inflation shot up to 7,20%, which suggests an aggressive increase in prices. For United States, inflation was a constant before COVID-19 pandemic, yet in 2020, inflation rose drastically up to an 8% due to the economic stimulus and supply chain disruptions. This drastic increase suggests a period of rapid price hikes across several sectors in the country's economy took place. For Panama, inflation maintained at a similar trajectory to Ecuador, having a negative inflation, or deflation, in 2019, and recovering by 2022, reaching a 2,86%. In Panama, inflation followed a trajectory like that of Ecuador, with negative inflation in 2019 and a recovery in 2022, reaching 2.86%, indicating a positive and moderate economic turnaround. It is possible to say that in general inflation rates in dollarized Latin American countries were stable before, during, and after the pandemic, following a similar trend. On the other hand, in the end of the pandemic (2022) those that experienced the greatest inflation were El Salvador and the USA.

Figure 3

Inflation Rates of Latin American Dollarized countries (Median), Colombia and Argentina.

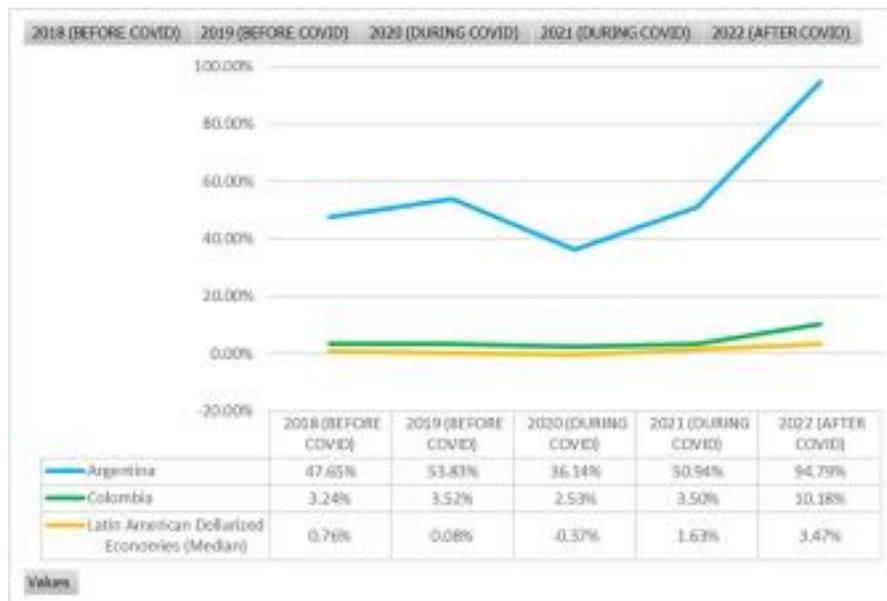

Note. Inflation rates of dollarized Latin American nations (median), Colombia, and Argentina. Data retrieved form (World Bank, 2023b)

In figure 3 it is possible to notice that before the pandemic (2018 & 2019) Argentina already had high inflation rates averaging over 50% in 2018, 2019, and remaining high for 2020. After the pandemic, Argentina experienced faster rise of inflation to 97.79%. In the case of Colombia with moderate inflation rates before the pandemic also saw a significant increase after the pandemic, from 3.50% to 10.18% in 2022. On the other hand, in the case of dollarized Latin American nation (median) experienced deflation of -0.37% during covid, while after the Covid-19 reached 3.47% which is high but much lower when compared to non-dollarized nations. Additionally, it is worth considering that inflation rated ted to increase in 2022 because covid vaccination program was a success, the society returned to their work, and at the same time as savings increased there was pressure over prices. In short, it is possible to say that apparently there exists a protection against inflation when being dollarized. Prove of that is that Latin American non-dollarized nations experienced greater inflation rates when compared to Latin American dollarized nations.

Figure 4

Gross Domestic Product Percentual Change (Absolut Value) in Latin American Dollarized Economies and the USA.

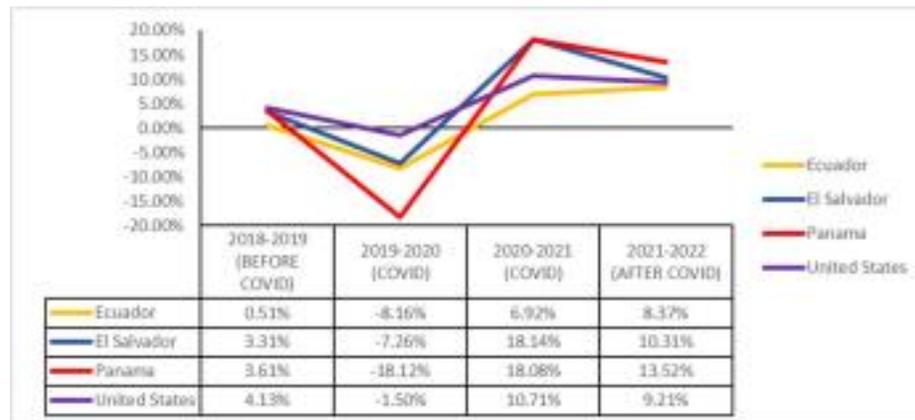

Note. Percentual change of the Gross Domestic Product of Latina American dollarized economies and The United States of America. Data retrieved from (World Bank, 2023a).

As a matter of fact, the Covid-19 pandemic affected every economy in the world due to several factors such as lockdowns, travel restrictions, bankruptcies, supply chain disruptions, and others. It is worth pointing out that for comparing Latin American dollarized economies, non-dollarized economies, and the US, GDP percentual changes will be considered instead of the whole GDP. This, with the purpose of simplifying the analysis that otherwise would be difficult due to the exponential economic differences that exist between the US and the rest. In figure 4, it is possible to have a look at how Latin American dollarized economies were doing before the start of the pandemic (2018-2019). In fact, Ecuador had a growth of 0.51 % that is small when compared to other countries, this happened because export grew slightly by 5.2%, households' consumption expenditures only rise by 1.5%, and the oil sector which is key for Ecuador's economic performance only grew 0.4% (Valencia, 2020). El Salvador grew 3.31%, because they have been working on developing its nation to the point where in 2019 they experienced a decline in the Gini index from 0.54 to 0.38 which means the economy is effectively distributing wealth throughout the society; in fact, they have been experiencing a decrease in poverty from 50% in 2011 to 28.8% in 2019 (Macrotrends, 2023a; World Bank, 2023g). Panama grew 3.61% this was due to Panama's plan for developing its country infrastructure and at the same time attracting new potential investors to the nation (Macrotrends, 2023b; Mandri-Perrott, 2016). Additionally, when talking about dollarization it is essential to consider the US as a reference because their economic measures most of the time have effects on dollarized nations as it was explained above. Before covid (2018-2019) the US economy grew 4.13%. As a matter of fact, once covid started (2019-2020), every economy suffered a contraction respectively, in the case of dollarized Latina American countries, Panama was the one who suffered the greatest contraction of -18.12%, this happened due to travel restrictions and the reduction of foreign investment from $4.22 billion to -$2.4 billion (Macrotrends, 2023c, 2023b). In the case of Ecuador, they suffered a contraction of -8.26% due to the country's dependence on oil, limited access to capital markets, and slow private sector (World Bank, 2023c). In the case of El Salvador, they suffered a contraction of -7.26% due to fiscal spending for contending the pandemic and for supporting households and businesses. Apart from that, their revenues were also severely affected due to supply chain disruptions (World Bank, 2023g). On the other hand, in the case of the USA, they experiment a contraction of only -1.50%, this happened due to their response which included unemployment insurance, stimulus checks, and support for small businesses apart from lowering interest rates to 0% (Furman & Wilson, 2021). Continuing the pandemic period (2020-2021), dollarized economies started to recover, in fact, El Salvador grew 18.14%, Panama grew 18.08%, and the USA grew 10.71% this type of growth was due to the accomplishment of the vaccination program in the mentioned countries. On the other hand, Ecuador only grew 6.92%, this happened due to the lack of confidence from international markets as well as its fiscal deficit (World Bank, 2023c). After Covid-19 (2021-2022), Ecuador grew 8.37%, El Salvador grew 10.31%, Panama grew 13,52%, and the USA grew 9.21%, this type of growth for each of the mentioned countries happened because the after Covid-19 social expenditure increase because citizen's from each of the mentioned countries had greater savings once the pandemic was over; hence, increasing consumer expenditure (World Bank, 2023c, 2023d, 2023g). In short, the behavior of dollarized Latin American nations during Covid-19 show that dollarization might provide some stability, it is not a cure for everything. The variety on economic recovery rates emphasize the importance of considering different factors that are not part of dollarization. In fact, the impressive recoveries of El Salvador and Panama emphasize the role that local variables play. In other words, a combination of consumer spending, governmental measures, access to financial markets, and so on are factors that determine the potential recovery that a country might have during disruption.

Figure 5

Gross Domestic Product Percentual Change (Absolute value) in Dollarized Latin American Countries (Median), Colombia, and Argentina.

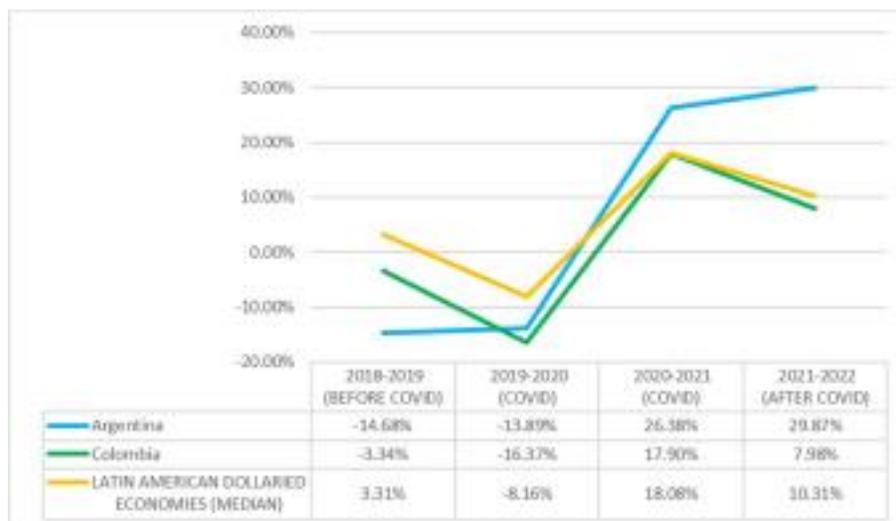

Note. Percentual changes of the Gross Domestic Product of Colombia, Argentina, and the median of dollarized Latin American economies (World Bank, 2023a).

As a matter of fact, in figure 5 before covid (2018-2019) Argentina experienced a contraction of -14.68%, this happened because of the inflation that Argentina was experiencing in 2019; in other words, the peso's devaluation and having a fiscal deficit increased the real value of Argentina's debt which is in dollars, and made it harder to access external financing (Nelson, 2019). Colombia in the same period experienced a contraction of -3.34%, this happened because of the fall of commodity prices such as oil industry and agricultural industry which accounts for 4.1% and 6.3% respectively of the Gross Domestic Product (Colombia Reports, 2023a; Oxford Business Group, 2019). On the other hand, the median growth rate of dollarized Latin American countries was 3.31%. When covid started (2019-2020) Colombia suffered a significant contraction of -16.37%, this happened because of the reduction of consumer expenditure in 15.1% , the negative effects that the pandemic had over the oil and mining industry that lead to a reduction of 15.7%, and the construction sector that face a contraction of 27.7% (Bocanegra et al., 2021). In the case of Argentina, their economy contracted -13.89%, this not only happened as a consequence of the Covid-19 pandemic; in fact, when the pandemic hit, they already were having problems to access lending markets; in other words, apart from the reduction of consumer expenditure, the reduction of international trade, the government had to print money in order to contra rest the effects of the pandemic (González, 2021). In the case of dollarized economies in Latin America the median contraction was -8.16%. Continuing the pandemic (2020-2021), Argentina grew 26.38%, this mainly because an increase in consumer expenditure due to the easing of the covid restrictions. Continuing the pandemic (2020-2021), Argentina grew 26.38%, this mainly because an increase in consumer expenditure due to the easing of the covid restrictions. In this case the main sectors that contribute to this growth were hotel and restaurants, transportation, and manufacturing (Buenos Aires Times, 2022; Reuters, 2022). In the case of Colombia, they experienced a growth of 17.9%, this happened as a result of easing pandemic restrictions; additionally, commodity prices started to rise, consumer expenditure increased (Medina, 2022). In the case of dollarized Latin American economies, they presented a median growth of 18.08%; however, it is worth considering that Ecuador did not grew that much due to the lack of confidence in international markets. Apart from that, all of them experiment a growth in different scales due to the success of the vaccination program and the increase of consumer expenditure. After Covid (2021-2022) the Argentinian Economy grew 29.87%, followed by Colombia with a growth of 7.98%, and Latin American dollarized economies with a median growth of 10.31%. These growths happened because the economy completely opened in 2022; hence, making it easier to trade and at the same time fomenting consumer expenditure (World Bank, 2023e, 2023f). In short, the behavior of Argentina, Colombia, and dollarized Latin American countries reveals different impacts. For example, pre-pandemic Argentina was struggling with inflation and debt while Colombia was severely affected by the drop of commodity prices. During the pandemic both countries had significant contractions due to the decrease of costumer spending and industry slowdowns. On the other hand, after the pandemic, Argentina presented a substantial growth that was greater than Colombia and dollarized economies. In other words, this suggests that dollarization might offer some economic stability, but it does not uniformly protect economies against economic disruptions. However, according to the former vice president of Ecuador, dollarization is not a substantial element for economic growth, in fact, by the difference in performance in dollarized Latin American nations. It is possible to say that economic measures and domestic policies play a bigger role when considering economic growth and development (Dahik, 2023).

Figure 6

Poverty Rate of Dollarized Latin American Countries and The USA.

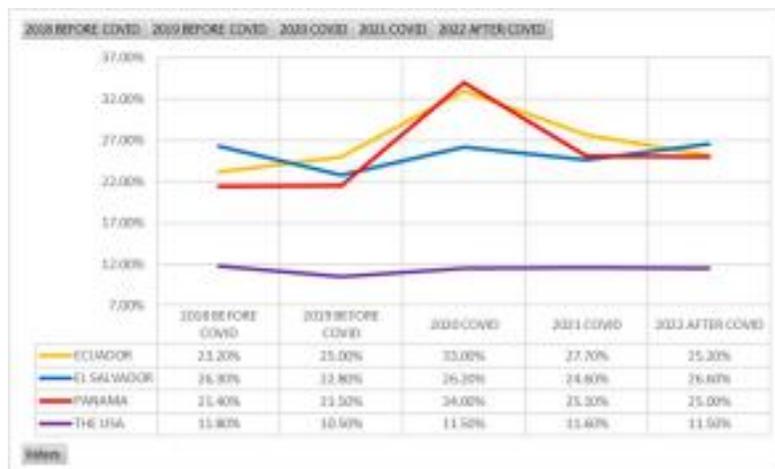

Note. Data retrieved from (Datosmacro, 2023b; INEC, 2023c; Panamá Sín Pobreza, 2022; Shrider & Creamer, 2023)

Notably, in figure 6 poverty rates are useful when evaluating the social environment of a country, especially when economic disruptions happen (Covid-19). In other words, analyzing poverty, it is possible to understand the society's economic situation and its conditions.

Considerably, the analysis of poverty rates from 2018 to 2022 for Dollarized Latin American countries and The USA reveals that each one of them experienced an increase of poverty during the pandemic. However, the impact was greater in Ecuador and Panama which could mean that they are more vulnerable when economic disruptions happen. On the other hand, The USA as it has a more stable economy and has the ability to modify its own monetary policy did not suffer that much; in fact, they had a peak of 11.50% due to their economic measures (Furman & Wilson, 2021). In other words, apparently factors other than dollarization such as Government response, levels of debt, access to financial markets, and economic structure play a critical role in economic resilience.

Figure 7

Poverty Rate of Dollarized Latin American Countries (Median), Colombia and Argentina.

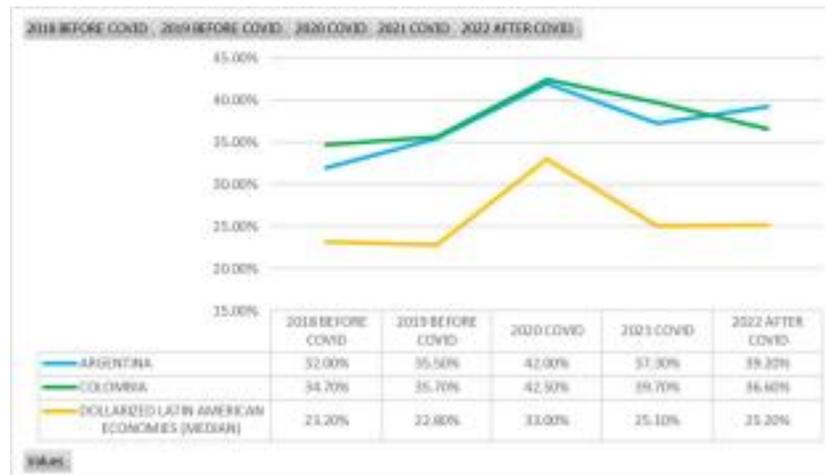

Note. Data retrieved from (Colombia Reports, 2023b; Datosmacro, 2023b, 2023a; INEC, 2023c; Panamá Sín Pobreza, 2022)

In figure 7 it is possible to notice that before the start of the pandemic (2018 & 2019) Argentina and Colombia had higher poverty rates. This suggests that economic differences and economic vulnerabilities is what lead to the difference in poverty rates between dollarized Latin American countries and non-dollarized Latin American countries. On the other hand, once the pandemic started, dollarized economies had a greater median move of 10.2% when compared to Colombia and Argentina which had a rise of 6.8% and 6.5% respectively. In other words, despite being dollarized, dollarized Latin American nations experienced a greater increase of poverty during the pandemic. On the other, after the pandemic poverty levels for dollarized countries increased by 2.4% and in the case of Argentina and Colombia, they increase by 3.7% and 0.9% respectively when compared to pre pandemic levels (2019). This clearly means that there is not a clear relationship when considering that dollarization serves as a protective mechanism against economic disruptions when talking about poverty. In other words, despite non dollarized nations having high poverty rates due to different factors apart from dollarization, when Covid-19 hit, dollarized nations and non-dollarized nation had a similar behavior; in fact, dollarized Latin American nations had a steeper increase in poverty when compared to non-dollarized economies in Latin America. On the other hand, it is worth considering that non-dollarized nations were having higher poverty rates before the pandemic.

## Discussion

Notably, a black swan is an event that cannot be predicted and when it happens it seems to have been inevitable (Britannica, 2023). In other words, Covid-19 pandemic was a black swan that could not be predicted and had a massive impact on the worldwide economy. As a matter of fact, this research is aiming to evaluate to what extent dollarization serves as a protective mechanism against economic disruptions such as Covid-19, for that reason, M1 and inflation, GDP, and poverty rate were analyzed with the purpose of assessing economic conditions as well as the social environment of each nation. In the case of inflation and M1 it is possible to say that there exists correlation between M1 and Inflation; in other words, this means that as governments increase the money supply it is excepted to experience inflation and therefore affecting the individual's purchasing power. During the analysis, it is possible to say that dollarization helped economies to cope better with inflation. In fact, the peak inflation in most of the countries happened after Covid-19 (2022) due to the rise of savings and the possibility of expending it. However, Latin American dollarized nations managed to have only a median inflation of 3.47% while Colombia and Argentina had 10.18% and 94.79% of inflation. This mainly happened due to not having the ability to employ monetary policy, especially given the history of being monetarily undisciplined. In the case of the gross domestic product, the analysis proves that despite being dollarized, they experienced almost similar effects as non-dollarized. In short, it seems that when considering economic performance, it is possible to say that dollarization does not necessarily serves as a protective mechanism; in fact, the nation's performance relies the most on domestic conditions and policies. In the case of poverty governmental response plays a bigger role; even so, it is important to consider that inflation levels influence poverty levels due to the reduction of the individual's purchasing power capacity. In this sense, dollarization might serve as a protective mechanism; nonetheless, a bigger role is played by governmental measures and policies. A clear example of that was the USA, by implementing its own monetary policy, they managed to not have a greater affection, especially because they have a record of being monetarily disciplined and properly managing the economy.

In this case, it is important to state that further research is needed to understand in what instances dollarization serves as a protective measure against economic disruption and to what extent governments participate in dollarization and if they take advantage of itself. In other words, to what extent the implementation of proper economic policies might be beneficial when combined with dollarization given the restriction of using monetary policy. On the other hand, it is worth pointing out that the limitation of this study consists of the lack of analysis of governmental action and its participation. In other words, through this study is possible to understand the macro environment of the considered economies; however, by understanding the role that domestic policies have over inflation, growth, and economic development. It will be possible to get a broader perspective on the extent to which dollarization serves as a protective mechanism against economic disruptions.

In conclusion after this analysis, is possible to say that dollarization might play an important role in protecting economies against high inflation rates; however, it is not clear to what extent they server as a protective mechanism against economic disruption when considering GDP and poverty rates. Given the history of dollarization, once it is implemented, inflation rates tend to decrease; nonetheless, in terms of economic performance and social conditions, it seems that domestic policies play a greater role. According to former vice president of Ecuador Alberto Dahik, dollarization in not a magic formula; instead, by having your own currency, countries have the ability to positively influence the economy. Nonetheless, the problem starts when fiscal undisciplined exists and when the government has not taken action to implement a proper virtuous circle that foments investment and productivity (Dahik, 2023). In short, monetary policy is an essential economic tool when properly used; however, given the history of monetary undisciplined in Latin American nations, by being dollarized, they are protected against their own monetary undisciplined. But, at the same without proper economic policies, dollarized countries are negatively impacting their own economy.

References

Álvarez, J. P. (2023, August 28). ¿Cómo le fue a Ecuador con la dolarización y qué esperar del plan Milei? Pros y contras. Bloomberg Línea.

https://www.bloomberglinea.com/latinoamerica/argentina/como-le-fue-a-

ecuador con-la-dolarizacion-y-que-esperar-del-plan-milei-pros-y-contras/

Archibold, R. C. (2013, March 23). A Once-Vibrant City Struggles as Panama Races

Ahead on a Wave of Prosperity. The New York Times.

https://www.nytimes.com/2013/03/24/world/americas/frustrations-in-colon-

panama as-economic-growth-skirts-by.html

Banco Central de la República de Argentina. (2023). Boletín Estadistico. Noviembre de

2023. Banco Central de Reserva de El Salvador. (n.d.). Oferta Estadística—Banco

Central de Reserva de El Salvador. Retrieved December 5, 2023, from

https://estadisticas.bcr.gob.sv/oferta-estadistica#21

Banco Central del Ecuador. (2023). Ecuador Interest Rate.

https://tradingeconomics.com/ecuador/interest-rate

Banco Central del Ecuador, B. C. del. (2000). Memoria Anual 1998 [Working Paper].

Quito: Banco Central del Ecuador, 2000. http://repositorio.bce.ec/handle/32000/2810

Banco de la República. (2023). Agregados monetarios | Banco de la República.

https://www.banrep.gov.co/es/estadisticas/agregados-monetarios

Basantes, X. (2020, May 2). Aportes y medidas frente a crisis han sido constantes en

40 años en Ecuador. El Comercio.

https://www.elcomercio.com/actualidad/negocios/aportes medidas-crisis-

historia-economia.html

Bocanegra, N., Vargas, C., & Symmes, J. (2021, February 15). Colombia economy

shrank 6.8% in 2020, in line with government forecast.

https://www.nasdaq.com/articles/colombia-economy-shrank-6.8-in-2020-in-line-

with government-forecast-2021-02-15

Booth, J., Wade, C., & Walker, T. (2018). Understanding Central America: Global

Forces, Rebellion, and Change. In Understanding Central America: Global Forces,

Rebellion, and Change (p. 374). https://doi.org/10.4324/9780429492624

Britannica. (2023, October 26). Black swan event | Definition, History, Examples, & Facts | Britannica. https://www.britannica.com/topic/black-swan-event

Brock, T. (2023, March 17). Monetary Policy Meaning, Types, and Tools. Investopedia. https://www.investopedia.com/terms/m/monetarypolicy.asp

Buenos Aires Times. (2022, February 25). GDP grew 10.3% in 2021 after three years of decline | Buenos Aires Times. Buenos Aires Times. https://www.batimes.com.ar/news/economy/indec-argentinas-gdp-grew-103-in-2021-following-three-years-of-decline.phtml

Ceja, L. G. (2016, March 8). The Banking Crisis That Nearly Destroyed Ecuador's Economy. https://www.telesurenglish.net/analysis/The-Banking-Crisis-That-Nearly-Destroyed Ecuadors-Economy-20160308-0047.html

Central Bank of Ecuador. (2023). ECONOMIC INFORMATION - Central Bank Of Ecuador. https://www.bce.fin.ec/en/economic-information

Central Reserve Bank of El Salvador. (2023). El Salvador Interest Rate. https://tradingeconomics.com/el-salvador/interest-rate

Chen, J. (2021, February 12). What Is Economic Collapse? Definition and How It Can Occur. Investopedia. https://www.investopedia.com/terms/e/economic-collapse.asp

CNN. (2023, November 10). Mahuad: La solución a una hiperinflación es la dolarización | Video. CNN. https://cnnespanol.cnn.com/video/hiperinflacion-dolarizacion-solucion modelo-argentina-conclusiones-tv/

Colantuoni, S. (2023, April 19). Dollarization in El Salvador celebrates 21 years. The Central American Group. https://www.thecentralamericangroup.com/dollarization-in-el salvador/

Colombia Reports. (2023a, January 25). Colombia's GDP statistics | Colombia Reports. Colombia News | Colombia Reports.

https://colombiareports.com/colombia-gdp statistics/

Colombia Reports. (2023b, September 25). Poverty and inequality | Colombia

Reports.  Colombia News | Colombia Reports.

https://colombiareports.com/colombia-poverty inequality-statistics/

Cortés Conde, R. (2003). LA CRISIS ARGENTINA DE 2001-2002. Cuadernos de

Economía, 40(121), 762–767. https://doi.org/10.4067/S0717-68212003012100049

Dahik, A. (2023, December 6). Opinions of Alberto Dahik, Former Ecuadorian Vice

President.

DANE, D. A. N. de E. (2023, September 7). Colombia Inflation Rate—August 2023

Data— 1955-2022 Historical—September Forecast.

https://tradingeconomics.com/colombia/inflation-cpi

Datosmacro. (2023a). Argentina—Riesgo de pobreza 2022 | Datosmacro.com.

https://datosmacro.expansion.com/demografia/riesgo-pobreza/argentina  Datosmacro.

(2023b). El Salvador—Riesgo de pobreza 2022 | Datosmacro.com.

https://datosmacro.expansion.com/demografia/riesgo-pobreza/el-salvador  Economia

y Desarrollo (Director). (2020, April 9). Política fiscal y crowding out: El caso de  la

CRISIS ARGENTINA | Cap. 38 - Macroeconomía.

https://www.youtube.com/watch?v=nuJgVdliZn0

EFE. (2022, April 21). Panamá, tercera economía con menor riesgo en América

Latina,  según JP Morgan. El Economista.

https://www.eleconomista.net/economia/Panama tercera-economia-con-menor-

riesgo-en-America-Latina-segun-JP-Morgan-20220421- 0024.html

El Universo. (2023, June 13). ¿Cómo afectó el fenómeno de El Niño de 1997 y 1998

al  Ecuador? El Universo. https://www.eluniverso.com/noticias/ecuador/como-

afecto-el fenomeno-de-el-nino-de-1997-y-1998-al-ecuador-nota/

Escobar, S. (2023, August 16). Dolarización: Cómo le fue a Ecuador 23 años

después de cambiar de moneda.

https://www.ambito.com/economia/dolarizacion-como-le-fue ecuador-23-

anos-despues-cambiar-moneda-n5796937

FED. (2023a). Federal Reserve Board—H6—Data Download Program—

Choose.

https://www.federalreserve.gov/datadownload/Choose.aspx?rel=H6

FED. (2023b). United States Fed Funds Interest Rate.

https://tradingeconomics.com/united states/interest-rate

Fernando, J. (2023, August 11). Inflation: What It Is, How It Can Be Controlled, and

Extreme  Examples. Investopedia. https://www.investopedia.com/terms/i/inflation.asp

Floyd, D. (2023). 10 Common Effects of Inflation. Investopedia.

https://www.investopedia.com/articles/insights/122016/9-common-

effects inflation.asp

Focus Economics. (2022, March 31). BanRep surprises and slows tightening of

conditions in  March. FocusEconomics. https://www.focus

economics.com/countries/colombia/news/monetary-policy/banrep-

surprises-and slows-tightening-of-conditions-in-march/

Furman, J., & Wilson, P. (2021, January 27). What the US GDP data tell us about

2020 |  PIIE. https://www.piie.com/blogs/realtime-economics/what-us-gdp-

data-tell-us about-2020

Galiani, S., Heymann, D., & Tommasi, M. (2003). Missed Expectations: The

Argentine  Convertibility. SSRN Electronic Journal.

https://doi.org/10.2139/ssrn.358380

García, N. (2013). La Crisis Financiera del Ecuador, 1998-2000. Economía y

Negocios, 4(1),  Article 1. https://doi.org/10.29019/eyn.v4i1.160

Gaspar, G. (2023, October 3). Inflación Argentina: Las Causas de un

Problema Macroeconomico. Noticias UNSAM.

https://noticias.unsam.edu.ar/2023/09/26/inflacion-argentina-las-causas-de-un problema-macroeconomico/

González, E. (2021, March 5). Argentina's perpetual crisis. EL PAÍS English. https://english.elpais.com/usa/2021-03-05/argentinas-perpetual-crisis.html  IMF. (2023). Panama Interest Rate. https://tradingeconomics.com/panama/interest-rate  INDEC: Instituto Nacional de Estadística y Censos de la República Argentina. (2023).  https://www.indec.gob.ar/indec/web/Nivel4-Tema-3-5-31

INEC. (2023a). Instituto Nacional de Estadística y Censo—Panamá.

https://www.inec.gob.pa/avance/Default2.aspx?ID_CATEGORIA=3&ID_CIFRAS=1 7&ID_IDIOMA=1

INEC. (2023b, September 6). Ecuador Inflation Rate—August 2023 Data—1970-2022 Historical—September Forecast. https://tradingeconomics.com/ecuador/inflation-cpi

INEC, I. N. de E. y C. (2023c). Encuesta Nacional de Empleo, Desempleo y Subempleo (ENEMDU), Diciembre 2022.

https://www.ecuadorencifras.gob.ec/documentos/web inec/POBREZA/2022/Diciembre_2022/202212_Boletin_pobreza.pdf

INEC, I. N. de E. y C. (2023d, June). Pobreza – junio 2023. Instituto Nacional de Estadística y Censos. https://www.ecuadorencifras.gob.ec/pobreza-por-ingresos/

Instituto Nacional de Estadística y Censos. (2023, July 1). Argentina Inflation Rate— September 2023 Data—1944-2022 Historical.

https://tradingeconomics.com/argentina/inflation-cpi

36Kelly, R. C., & Costagliola, D. (2021, March 9). Monetary Aggregates: Definition and Examples. Investopedia.

https://www.investopedia.com/terms/m/monetary aggregates.asp

Kopar, L. A. (2018). The Macroeconomics of Dollarization: A Cross-Country Examination of the Effects of Dollarization in Ecuador and El Salvador.

Macrotrends. (2023a). El Salvador Poverty Rate 1989-2023.

https://www.macrotrends.net/countries/slv/el-salvador/poverty-rate

Macrotrends. (2023b). Panama Foreign Direct Investment 1977-2023.

https://www.macrotrends.net/countries/PAN/panama/foreign-direct-investment

Macrotrends. (2023c). Panama Tourism Statistics 1995-2023.

https://www.macrotrends.net/countries/PAN/panama/tourism-

statistics  Macrotrends. (2023d). World Inflation Rate 1981-2023.

https://www.macrotrends.net/countries/WLD/world/inflation-rate-cpi

Mandri-Perrott, C. (2016, August 24). Prioritizing infrastructure investments:

Panama's  long-term path to PPPs.

https://blogs.worldbank.org/ppps/prioritizing-infrastructure investments-

panama-s-long-term-path-ppps

McKitrick, R. (2022, May 22). Inflation—Why now and not post-2008?: Op-ed. Fraser

Institute. https://www.fraserinstitute.org/article/inflation-why-now-and-not-post-2008

Medina, O. (2022, February 15). Record expansion: Colombia's economy grows the

most in  115 years. Al Jazeera. https://www.aljazeera.com/economy/2022/2/15/record

expansion-colombias-economy-grows-the-most-in-115-years

Meléndez, Á. (2022, September 20). EXCLUSIVA: Jamil Mahuad y por qué se

dolarizó al  Ecuador. Bloomberg Línea.

https://www.bloomberglinea.com/2022/09/20/exclusiva jamil-mahuad-y-por-

que-se-dolarizo-al-ecuador/

National Bureau of Economic Research. (2023, September 9). Unpacking the

Causes of  Pandemic-Era Inflation in the US. NBER.

https://www.nber.org/digest/20239/unpacking-causes-pandemic-era-

inflation-us  Nelson, R. M. (2019, October 10). Argentina's Economic Crisis.

WITA.  https://www.wita.org/atp-research/argentinas-economic-crisis/

Ortiz, D. (2023, January 11). 5 beneficios de la dolarización en Ecuador, tras 23 años. El Comercio. https://www.elcomercio.com/actualidad/negocios/dolarizacion-ecuador aniversario-enero-2023.html

Oxford Business Group. (2019, July 8). Colombia's economy set for expansion despite challenges—Colombia 2019—Oxford Business Group. https://oxfordbusinessgroup.com/reports/colombia/2019-report/economy/small-steps moderate-growth-is-on-the-horizon-amid-strain-from-migrant-populations-and subdued-commodity-prices

Panamá Sín Pobreza. (2022, October 19). La Pobreza en Panamá 2023. Panorama general: Pobreza, pobreza extrema e indigencia por ingreso. Iniciativa Panamá Sin Pobreza. https://panamasinpobreza.org/panama/pobreza/

Patiño, O. (2023, November 13). Opinions From An Expert In Economics & International Finance.

Pettinger, T. (2022, November 16). To what extent did Covid cause inflation? Economics Help. https://www.economicshelp.org/blog/169072/economics/to-what-extent-did covid-cause-inflation/

Rath, L. (2022, December 31). Coronavirus History: Origin and Evolution. WebMD. https://www.webmd.com/covid/coronavirus-history

Reed, L. W. (2020, January 7). Celebrating Ecuador's Dollarization | Lawrence W. Reed. https://fee.org/articles/celebrating-ecuador-s-dollarization/

Reuters. (2022, March 23). Argentina's GDP rose 10.3%, the biggest advance in 17 years. LABS English. https://labsnews.com/en/news/economy/argentina-gdp-2021/

Roura, A. (2020, January 9). ¿Por qué sigue siendo tan popular la dolarización entre los ecuatorianos? (Y cuál es su lado oscuro). BBC News Mundo.

https://www.bbc.com/mundo/noticias-america-latina-50916554

Santayana, G. (1905). The Life of Reason: Reason in Common Sense (1–5).

Shrider, E. A., & Creamer, J. (2023). Poverty in the United States: 2022.  Suarez, J. (2022, July 13). How the Federal Reserve uses expansionary monetary policy to stimulate growth during an economic downturn. Business Insider.

https://www.businessinsider.com/personal-finance/what-is-expansionary-monetary policy

Swiston, A. (2011). Official Dollarization as a Monetary Regime: Its Effects on El Salvador.  Swiston, A. (2012). Chapter 7: Official Dollarization in El Salvador as an Alternative  Monetary Framework. In Central America, Panama, and the Dominican Republic.  International Monetary Fund.

https://www.elibrary.imf.org/display/book/9781616353780/ch007.xml  Taylor, S. (2020, February 18). Quantitative Analysis. Corporate Finance Institute.

https://corporatefinanceinstitute.com/resources/data-science/quantitative-analysis/

Tech Target. (n.d.). What is Qualitative Data? CIO. Retrieved October 19, 2023, from  https://www.techtarget.com/searchcio/definition/qualitative-data

United Nations. (2016, May). Economic Recovery after Natural Disasters. United Nations;  United Nations. https://www.un.org/en/chronicle/article/economic-recovery-after natural-disasters

Valencia, A. (2020, April 1). Economía de Ecuador creció 0,1% en 2019 impulsada por  exportaciones: Banco central. Reuters. https://www.reuters.com/article/idUSKBN21J5MB/

Vasquez, J. (2023, November). Opinions of an Expert in Finance.

World Bank. (n.d.-a). GDP growth (annual %)—Ecuador. World Bank Open Data. Retrieved  November 20, 2023, from https://data.worldbank.org/indicator/NY.GDP.MKTP.KD.ZG?end=2022&locatio

ns= EC&start=1996

World Bank. (n.d.-b). Inflation, Consumer Prices (annual %)—Ecuador. World Bank

    Open Data. Retrieved November 20, 2023, from https://data.worldbank.org

World Bank. (2023a). GDP [Excel].

    https://data.worldbank.org/indicator/NY.GDP.MKTP.CD

World Bank. (2022). WDR 2022 Chapter 1. Introduction [Text/HTML]. World Bank.

    https://www.worldbank.org/en/publication/wdr2022/brief/chapter-1-introduction-

    the economic-impacts-of-the-covid-19-crisis

World Bank. (2023b). Inflation, consumer prices (annual %). World Bank Open

    Data. https://data.worldbank.org

World Bank. (2023c). Overview Ecuador [Text/HTML]. World Bank.

    https://www.worldbank.org/en/country/ecuador/overview

World Bank. (2023d). Overview Panama [Text/HTML]. World Bank.

    https://www.worldbank.org/en/country/panama/overview

World Bank. (2023e, June 7). Overview Colombia [Text/HTML]. World

    Bank. https://www.worldbank.org/en/country/colombia/overview

World Bank. (2023f, October 2). Overview Argentina [Text/HTML]. World

    Bank. https://www.worldbank.org/en/country/argentina/overview

World Bank. (2023g, October 4). Overview El Salvador [Text/HTML]. World

    Bank. https://www.worldbank.org/en/country/elsalvador/overview

World Economic Forum. (2022, July 4). Interest rate hikes vs inflation: How are

    different countries doing it? World Economic Forum.

    https://www.weforum.org/agenda/2022/07/interest-rate-hikes-inflation-rate-economy/

Annexes

Annex 1

Interviews Question

Former Vice-president

- At that time, what were your thoughts about dollarization given the economic turmoil that the nation was going through.
- What were the biggest challenges Ecuador faced during the dollarization transition?
- How has dollarization impacted inflation and economic growth in Ecuador?
- What is your opinion with respect to dollarization, was it convenient then? Is it still convenient now?
- Do you consider that during the pandemic, dollarization played an important role when providing economic stability?

Professor's Interview

- What do you consider to be the main benefits and drawbacks of dollarization for Ecuador?
- What determines the success of dollarization?

- What do you consider to be challenges and opportunities for Ecuador as a dollarized country in the post-pandemic era?